\documentclass[3p,times,twocolumn]{elsarticle}

\usepackage{ecrc}
\usepackage{amssymb,amsmath,amsthm,bm}

\volume{00}

\firstpage{1}

\journalname{Nuclear Physics B Proceedings Supplement}

\runauth{K.A. Kouzakov and A.I. Studenikin}

\jid{nuphbp}

\jnltitlelogo{Nuclear Physics B Proceedings Supplement}

\usepackage{amssymb}

\biboptions{sort&compress}

\usepackage[figuresright]{rotating}

\begin{document}

\begin{frontmatter}



\dochead{}

\title{Neutrino magnetic moment, millicharge and charge radius}

\author[1]{Konstantin A. Kouzakov}
\ead{kouzakov@srd.sinp.msu.ru}
\author[2,3]{Alexander I. Studenikin}
\ead{studenik@srd.sinp.msu.ru}
\address[1]{Department of Nuclear Physics and Quantum Theory of Collisions, Faculty of Physics, Lomonosov Moscow State University, 119991 Moscow , Russia}
\address[2]{Department of Theoretical Physics, Faculty of Physics, Lomonosov Moscow State University, 119991 Moscow , Russia}
\address[3]{Joint Institute for Nuclear Research, 141980 Dubna , Moscow Region, Russia}
\begin{abstract}
A brief overview of neutrino electromagnetic properties is presented from a theoretical perspective. Their potential effects on coherent elastic neutrino-nucleus scattering are outlined.
\end{abstract}

\end{frontmatter}



\section{Introduction}


Within the original formulation of the Standard Model (SM), neutrinos appear to be massless particles. However, currently it is commonly accepted that the SM should be extended to some more general theory, in particular, because of neutrinos which are the only particles exhibiting experimentally properties beyond the SM (BSM). The latter include neutrino mixing and oscillations supported by the discovery of flavour conversion of neutrinos from different sources, that is, the effect which is not possible for massless neutrinos.

In many SM extensions, which account for neutrino masses and mixing, neutrinos acquire nontrivial electromagnetic properties, thus allowing direct neutrino interactions with electromagnetic fields and charged particles or with particles which have magnetic moments. Unfortunately, in spite of much effort, up to now no experimental evidence has been found, neither from terrestrial laboratory measurements nor from astrophysical observations, in favor of nonvanishing neutrino electromagnetic characteristics. At the same time, once these characteristics are experimentally confirmed, they will open a door to the BSM physics.

The importance of neutrino electromagnetic characteristics was first mentioned by Wolfgang Pauli in 1930, when he postulated the existence of neutrino and supposed that its mass can be of the order of that of the electron. In his famous letter to ``radiative ladies and gentlemen'' Pauli also discussed the possibility that neutrino could posses a magnetic moment. It is worth to mention another early paper \cite{Touschek:1957} that concerns the magnetic moment of neutrino. The authors showed that in the zero-mass limit the neutrino magnetic moment also should tend to zero.

Systematic theoretical studies of neutrino electromagnetic interactions have started shortly after it was shown that in the minimally-extended SM with right-handed neutrinos the magnetic moment of a massive neutrino is, in general, non-zero and that its value is determined by the neutrino mass \cite{Fujikawa:1980yx}. For the recent reviews on the neutrino electromagnetic properties and related problems see \cite{Studenikin:2008bd, Giunti:2008ve, Broggini:2012df, Kouzakov:2014, Giunti:2014ixa}. In this contribution we give a very brief overview of the neutrino magnetic moment, millicharge and charge radius and discuss how they can manifest themselves in neutrino-nuclues coherent scattering.

\section{Neutrino electromagnetic vertex}

The neutrino electromagnetic properties are determined by the neutrino electromagnetic vertex
function $\Lambda_{\mu}(q)$, which is related to the matrix element of the electromagnetic current $J_{\mu}^{EM}$ between the
neutrino initial $\psi_i(k)$ and final $\psi_j(k')$ mass states, $p^2=m^2_i$ and
$p'^2=m^2_j$,
\begin{equation}\label{matr_elem}
\langle{\psi}_j(k^{\prime})|J_{\mu}^{EM}|\psi_i(k)\rangle= {\bar
u}_j(k^{\prime})\Lambda_{\mu}(q)u_i(k).
\end{equation}
In the most general case consistent with Lorentz and electromagnetic gauge
invariance \cite{Kayser:1982br,Nieves:1981zt,
Nowakowski:2004cv,Giunti:2014ixa} the electromagnetic vertex function can be presented in the form
\begin{eqnarray}
\label{Lambda} \Lambda_{\mu}(q)_{ij}=
f_{Q}(q^{2})_{ij}\gamma_\mu-f_{A}(q^{2})_{ij}
(q^{2}\gamma_{\mu}-q_{\mu}\! \! \not{\! q})\gamma_{5}\nonumber
\\ - f_{M}(q^{2})_{ij}\sigma_{\mu\nu}q^{\nu}
+if_{E}(q^{2})_{ij}\sigma_{\mu\nu}q^{\nu}\gamma_{5},
\end{eqnarray}
where
$f_{Q} $, $f_{A}$, $f_{M} $ and $f_{E} $ are respectively charge, anapole, dipole magnetic and electric electromagnetic form factors, and $\sigma_{\mu\nu}=(\gamma _{\mu}\gamma_{\nu}-\gamma _{\nu}\gamma_{\mu})/2$. The form factors depend on the Lorentz invariant dynamical quantity $q^2$, with $q=k-k'$ being the photon four-momentum. It should be also noted that the properties of the form factors are quite different for the Dirac and Majorana neutrinos (see \cite{Giunti:2014ixa} and references therein).

 From the demand that the form factors at
zero momentum transfer, $q^2=0$, are elements of the scattering
matrix, it follows that  in any consistent theoretical model the form
factors in the matrix element (\ref{matr_elem}) should be gauge-independent and finite. Then, the form factors values at $q^{2}=0$
determine the static electromagnetic properties of the neutrino that
can be probed or measured in the direct interaction with external
electromagnetic fields.

It is usually believed that the
neutrino electric charge $e_\nu=f_Q(0)$ is zero. This is often thought to be
attributed to the gauge-invariance and anomaly-cancellation constraints
imposed in the SM. In the SM of $SU(2)_L
\times U(1)_Y$ electroweak interactions it is possible to get
\cite{Foot:1992ui} a general proof that neutrinos are
electrically neutral, which is based on the requirement of electric
charges' quantization. The direct calculations of the neutrino charge
in the SM for massless (see, for instance
\cite{Bardeen:1972vi,CabralRosetti:1999ad})
and massive neutrinos \cite{Dvornikov:2003js,Dvornikov:2004sj}  also prove that,
at least at the one-loop level, the neutrino electric charge is
gauge-independent and vanishes. However, if the neutrino has a mass,
it still may become electrically millicharged. A brief discussion of different mechanisms for introducing millicharged particles including neutrinos can be found in \cite{Davidson:2000hf}. It should be mentioned that the most stringent experimental constraints on the electric charge of the neutrino can be obtained from
neutrality of matter. This yields $e_\nu \lesssim 10^{-21} e_0$ \cite{Bressi:2011pj}. The most stringent astrophysical constraint $e_\nu \lesssim 1.3 \times 10^{-19} e_0$ has been obtained recently \cite{Studenikin:2012vi}. A detailed discussion on
other constraints on $e_\nu$  can be found in \cite{Giunti:2014ixa}.

Even if the electric charge of a neutrino is zero, the electric
form factor $f_Q(q^2)$ can still contain nontrivial information about
neutrino static properties \cite{Giunti:2014ixa}. A neutral particle can be characterized
by a superposition of two charge distributions of opposite signs, so
that the particle form factor $f_Q(q^2)$ can be non-zero for
$q^2\neq 0$. The mean charge radius (in fact, it is the charged
radius squared) of an electrically neutral neutrino is given by
\begin{equation}\label{nu_cha_rad}
{\langle r_{\nu}^2\rangle}={6}\left.\frac{df_{Q}(q^2)}{dq^2}\right|_{ q^2=0},
\end{equation}
which is determined by the second term in the power-series expansion of the
neutrino charge form factor.

Note that there is a long-standing discussion (see \cite{Giunti:2014ixa}
for details) on the possibility to obtain
 for the neutrino charged radius a gauge-independent and
finite quantity. In the corresponding calculations, performed in the
one-loop approximation including additional terms from the $\gamma-Z$
boson mixing and the box diagrams involving $W$ and $Z$ bosons, the
following gauge-invariant result for the neutrino charge radius have
been obtained \cite{Bernabeu:2004jr}:
${\langle r_{\nu_e}^2\rangle}=4
\times 10^{-33}$\,cm$^2$. This theoretical result differs at most
by an order of magnitude from the available experimental bounds on
$\langle r_{\nu_i}^2\rangle$ (see \cite{Giunti:2014ixa} for references and more detailed
discussion). Therefore, one may expect that the experimental
accuracy will soon reach the level needed to probe the neutrino
effective charge radius.

The most well studied and understood among the neutrino electromagnetic characteristics are the dipole magnetic and electric moments, which are given by the corresponding form factors at $q^2=0$:
\begin{equation}
\mu_{ij}=f_{M}(0)_{ij}, \qquad \epsilon_{ij}=f_{E}(0)_{ij}.
\end{equation}
The diagonal magnetic and electric moments of a Dirac neutrino in the minimally-extended SM with right-handed neutrinos, derived for the first time in \cite{Fujikawa:1980yx}, are respectively
\begin{equation}\label{mu_D}
    \mu^{D}_{ii}
  = \frac{3e_0 G_F m_{i}}{8\sqrt {2} \pi ^2}\approx 3.2\times 10^{-19}
  \mu_{B}\left(\frac{m_i}{1 \, \text{eV}}\right), \qquad \epsilon^{D}_{ij}=0,
  \end{equation}

where $\mu_B$ is the Borh magneton.
 According to (\ref{mu_D}) the value of the neutrino magnetic moment is very small.
However, in many other theoretical frameworks (beyond the minimally-extended SM) the neutrino magnetic moment can reach values that are of interest for the next generation of terrestrial experiments and also accessible for astrophysical observations.Note that the best laboratory upper limit on a neutrino magnetic moment, $\mu_{\nu} \leq 2.9 \times 10^{-11} \mu_{B}$, has been obtained by the GEMMA collaboration  \cite{Beda:2012zz}, and the best astrophysical limit is  $\mu_{\nu}\leq 3 \times
10^{-12} \mu _B$ \cite{Raffelt:1990pj}.

\section{Neutrino-nucleus coherent scattering}
Currently, the most sensitive
probe of neutrino electromagnetic properties
is provided by direct laboratory measurements of
(anti)neutrino-electron scattering at low energies in solar,
accelerator and reactor experiments (their detailed description can be found in~\cite{Wong:2005pa, Balantekin:2006sw, Beda:2007hf,Giunti:2008ve, Broggini:2012df,Giunti:2014ixa}). Below we focus on the coherent elastic neutrino-nucleus scattering~\cite{Freedman:1974}, which has not been experimentally observed so far, but which is expected to be accessible in the reactor experiments when lowering the energy threshold of the employed Ge detectors~\cite{Wong:2011,Li:2013,Li:2014}.

We consider the case of neutrino scattering off a spin-zero nucleus with even numbers of protons and neutrons, $Z$ and $N$. The matrix element of this process, taking into account the neutrino electromagnetic properties, reads
\begin{eqnarray}
\label{M}
\mathcal{M}=\left[\frac{G_F}{\sqrt{2}}\bar{u}(k^\prime)\gamma^\mu(1-\gamma_5)u(k)C_V\right.\nonumber\\
+\frac{4\pi Ze_0}{q^2}\left({e_\nu}+\frac{e_0}{6}q^2\langle r_\nu^2\rangle\right)\bar{u}(k^\prime)\gamma^\mu u(k)\nonumber\\
\left.-\frac{4\pi Ze_0\mu_\nu}{q^2}\bar{u}(k^\prime)\sigma^{\mu\nu}q_\nu u(k)\right]\mathcal{J}_\mu,
\end{eqnarray}
where $C_V=[Z(1-4\sin^2\theta_W)-N]/2$, $\mathcal{J}_\mu=(p_\mu+p_\mu^\prime)F(q^2)$, with $p$ and $p^\prime$ being the initial and final nuclear four-momenta, and $\mu_\nu$, $e_\nu$, and $\langle r_\nu^2\rangle$ are the neutrino effective magnetic moment, millicharge, and charge radius squared, respectively.
For neutrinos with energies of a few MeV the maximum momentum transfer squared ($|q^2|_{\rm max}=4E_\nu^2$) is still small compared to $1/R^2$, where $R$, the nucleus radius, is of the order of $10^{-2}-10^{-1}$\,MeV$^{-1}$. Therefore, the nuclear elastic form factor $F(q^2)$ can be set equal to one. Using (\ref{M}), one obtains the differential in the energy transfer $T$ cross section as a sum of two components.
The first component conserves the neutrino helicity and can be presented in the form
\begin{equation}
\label{h.-c.}
\frac{d\sigma_{1}}{dT}=\eta^2\,\frac{d\sigma_{SM}}{dT},
\end{equation}
where
$$
\eta=1-\frac{\sqrt{2}\pi e_0Z}{G_FC_V}\left[\frac{e_\nu}{MT}-\frac{e_0}{3}\langle r_\nu^2\rangle\right],
$$
with $M$ being the nuclear mass, and
\begin{equation}
\label{SM}
\frac{d\sigma_{SM}}{dT}=\frac{G_F^2M}{\pi}\left(1-\frac{T}{2E_\nu}-\frac{MT}{2E_\nu^2}\right)C_V^2
\end{equation}
is the SM cross section due to weak interaction~\cite{Drukier:1984}. The second, helicity-flipping component is due to the magnetic moment only and is given by~\cite{Vogel:1989}
\begin{equation}
\label{NMM}
\frac{d\sigma_{2}}{dT}=4\pi e_0^2\mu_\nu^2\,\frac{Z^2}{T}\left(1-\frac{T}{E_\nu}+\frac{T^2}{4E_\nu^2}\right).
\end{equation}

Clearly, any deviation of the measured cross section of the discussed process from the very precisely known SM value~(\ref{SM}) will provide a signature of the BSM physics (see also~\cite{Scholberg:2006,Barranco:2005,Barranco:2007,Davidson:2003}). Formulas~(\ref{h.-c.}) and~(\ref{NMM}) describe such a deviation due to neutrino electromagnetic interactions.

%
\section*{Acknowledgments}
One of the authors (A.I.S.) is thankful to Paolo Bernardini, Gianluigi Fogli and
Eligio Lisi for the invitation to attend the Neutrino Oscillation
Workshop and to all of the organizers for their kind hospitality in
Conca Specchiulla. This study has been partially supported by the Russian Foundation for Basic Research (Grants no. 14-01-00420-a, no. 14-22-03043-ofi-m, and no. 15-52-53112-gfen-a).











\end{document}